\documentclass[reprint, amsmath,amssymb, aps, onecolumn]{revtex4-2}
\usepackage[utf8]{inputenc}
\usepackage{graphicx}
\usepackage{bm}
\usepackage{xcolor}

\usepackage{here}

\usepackage{booktabs}
\usepackage{tabularx}
\usepackage{array}
\newcolumntype{T}[1]{>{\ttfamily\footnotesize\raggedright\arraybackslash}p{#1}}

\usepackage{algorithm}
\usepackage{algpseudocode}
\usepackage{caption}

\usepackage[colorlinks=true, linkcolor=blue, citecolor=blue, urlcolor=blue]{hyperref}

\usepackage{doi}

\AtBeginDocument{%
  \algrenewcommand\alglinenumber[1]{\scriptsize #1}%
}

\usepackage{listings}
\lstdefinestyle{promptstyle}{
  basicstyle=\ttfamily\footnotesize,
  columns=fullflexible,
  keepspaces=true,
  breaklines=true,
  breakatwhitespace=false,
  breakindent=0pt,
  breakautoindent=false,
  xleftmargin=0pt,
  showstringspaces=false,
  showspaces=false,
  showtabs=false
}

\usepackage{fvextra}
\fvset{
  fontsize=\small,
  breaklines=true,
  breakanywhere=true,
  breaksymbolleft={},   
  breaksymbolright={},  
  breaksymbol={}        
}

\begin{document}

\title{LOGOS-CA: A Cellular Automaton Using Natural Language as State and Rule}
\author{Keishu Utimula}
\date{\today}

\begin{abstract}

Large Language Models (LLMs), trained solely on massive text data, 
have achieved high performance on the Winograd Schema Challenge (WSC), 
a benchmark proposed to measure commonsense knowledge and reasoning abilities about the real world.
This suggests that the language produced by humanity describes a significant portion of the world with considerable nuance.
In this study, we attempt to harness the high expressive power of language within cellular automata.
Specifically, we express cell states and rules in natural language and delegate their updates to an LLM.
Through this approach, cellular automata can transcend the constraints of merely numerical states and fixed rules, 
providing us with a richer platform for simulation.
Here, we propose LOGOS-CA (Language Oriented Grid Of Statements – Cellular Automaton) as a natural framework to achieve this and examine its capabilities.
We confirmed that LOGOS-CA successfully performs simple forest fire simulations 
and also serves as an intriguing subject for investigation from an Artificial Life (ALife) perspective.
In this paper, we report the results of these experiments and discuss directions for future research using LOGOS-CA.

\end{abstract}

\maketitle

\section{Introduction}
\label{sec.intro}

For every known phenomenon, attempts have been made to describe it through ``language.''
Here, let us take this ``language'' to include numbers, symbols, and the mathematical equations that combine them.
In \textit{Course in General Linguistics}, Saussure argued that language is not merely a label 
but an arbitrary association between a sound-image and a concept, 
and further suggested a view of language as a system that carves out phenomena~\cite{1959SAU}.
On the other hand, the early Wittgenstein stated that 
the world is the totality of facts and that propositions are pictures of facts~\cite{1922WIT}, 
while in his later work he positioned the act of describing the world as one form of ``language-games.''~\cite{1953WIT}
It can be argued that humanity has refined this craft each time it encounters unknown phenomena.

\vspace{2mm}
Prior to the emergence of Large Language Models (LLMs), in the context of classical symbolic AI, 
it was widely believed that an explicit world model was required for AI to understand the real world~\cite{1969MCC, 2001MES}. 
The Winograd Schema Challenge (WSC) is one of the most representative benchmarks proposed to measure commonsense knowledge and reasoning abilities about the real world~\cite{2011LEV}. 
The authors designed it as a problem that cannot be solved simply through a large corpus and statistical processing on it, and it has been used to investigate AI's understanding of the real world~\cite{2011LEV, 2020SAK}. 
However, Trinh and Le showed that an RNN-based language model trained on massive text data in an unsupervised manner achieved $63.7$\% accuracy on the WSC~\cite{2018TRI}.
Subsequently, GPT-2 improved its score within the same framework, and the following fine-tuned BERT achieved $90$\% accuracy~\cite{2019RAD, 2020SAK, 2023KOC}.
These results do not immediately negate the necessity of a world model for the WSC; 
indeed, some argue that an implicit world model is constructed internally within LLMs~\cite{2019PET, 2024VAF, 2025XIE}.
In any case, these results suggest that the language humanity has produced captures a significant portion of the world with considerable nuance.

\vspace{2mm}
The expressive power of language is unfathomable.

\vspace{2mm}
Here, we consider combining the high expressive power of language with cellular automata.

\vspace{2mm}
A cellular automaton is a discrete computational model defined by a lattice of cells, each holding a state, and rules governing their updates~\cite{2011KAR}.
This was introduced by John von Neumann as a theoretical framework for discussing self-replicating machines~\cite{1966NEU, 1986WOL, 1998SIP, 2004SAL}. 
Typically, a cell's state is one of several integers, and the rules are expressed as program code.

\vspace{2mm}
Cellular automata have been used for dynamical system computations including fluid dynamics~\cite{1986FRI, 1988ROT, 2005WOL}, 
simulations of natural phenomena such as forest fires~\cite{1971KOU, 1990BAK, 1992DRO}, 
analysis of macroscopic phenomena such as traffic congestion and urban development~\cite{1992NAG, 2022CHA}, 
and even certain types of cryptography~\cite{1994NAN}.
In such applications, cell states are no longer limited to integers; they may also be real numbers representing physical quantities~\cite{2001RAU, 2019CHA, 2020MOR, 2022MOR}.

\vspace{2mm}
Now, what would happen if we expressed both cell states and their update rules in natural language?
This can be achieved by expressing cell states and rules in natural language and delegating their updates to an LLM. 
This makes it possible to bring the high expressive power of natural language into the realm of simulation.

\vspace{2mm}
In this paper, we propose LOGOS-CA (Language Oriented Grid Of Statements – Cellular Automaton) as a natural implementation of this.
Here, both cell states and their update rules are described in natural language within each cell. 
The simulation proceeds by having an LLM, given the descriptions of the target cell and its Moore neighborhood, 
return the next description of the target cell.
To experimentally validate this framework, we conducted simulations of forest fires and Artificial Life (ALife). 
First, by running a simple forest fire simulation on LOGOS-CA, we verify that LOGOS-CA can execute cellular automata based on simple rules.
Next, we run an ALife simulation for the purpose of testing freer descriptions.
Finally, based on these results, we discuss simulations using LOGOS-CA.

\vspace{2mm}
\section{Model and Methodology}
\label{sec.method}

The framework of LOGOS-CA proposed in this study is remarkably simple.
Specifically, there is a two-dimensional plane composed of cells, 
each possessing some description, and an LLM determines 
the next description of a given cell based on the descriptions of that cell and its Moore neighborhood.
The pseudocode is presented in Algorithms~\ref{alg:nlca_compact} and~\ref{alg:next_compact}.
Here, the prompts given to the LLM are provided in Table~\ref{tab:prompt-templates}.
The system prompt remains fixed, and cell information is provided to the LLM through the user prompt.
\rm{\{target\_cell\}} and \rm{\{neighbor\_desc\}} are replaced with 
the description of the current target cell and a structured string of its current Moore neighborhood descriptions 
with position labels \rm{(top-left, top, ..., bottom-right)} formatted as ``\rm{- <position>: <description>}'', respectively.

\vspace{2mm}
A key feature of LOGOS-CA is that it enables various simulations simply by changing the initial cell descriptions, without modifying how prompts are provided.

\begin{center}
\begin{minipage}[t]{0.48\linewidth}
\captionsetup{type=algorithm}
\captionof{algorithm}{LOGOS-CA with an LLM}
\label{alg:nlca_compact}
\hrule\vspace{2mm}
\begin{algorithmic}[1]
\Require size $(X,Y)$, steps $T$, model $M$
\Ensure history $\mathcal{H}=\{G_0,\dots,G_T\}$
\State $G_0 \gets$ \textsc{Init}$(X,Y)$
\State $\mathcal{H} \gets [G_0]$
\For{$t=0$ to $T-1$}
  \State $G_{t+1} \gets$ empty grid
  \For{$i=0$ to $Y-1$}
    \For{$j=0$ to $X-1$}
      \State $\mathcal{N}\gets$ \textsc{Nbr}$(G_t,i,j)$
      \State $G_{t+1}[i,j]\gets$ \textsc{Next}$(M,G_t[i,j],\mathcal{N})$
    \EndFor
  \EndFor
  \State append $G_{t+1}$ to $\mathcal{H}$
\EndFor
\State \Return $\mathcal{H}$
\end{algorithmic}
\vspace{2mm}\hrule
\end{minipage}\hfill
\begin{minipage}[t]{0.48\linewidth}
\captionsetup{type=algorithm}
\caption{\textsc{Next}: one-cell update}
\label{alg:next_compact}
\hrule\vspace{2mm}
\begin{algorithmic}[1]
\Require $M$, state $s$, neighbors $\mathcal{N}$
\Ensure next state $s'$
\State $p\gets$ \textsc{Prompt}$(s,\mathcal{N})$
\State $r\gets$ \textsc{CallLLM}$(M,p)$
\State $j\gets$ \textsc{ParseJSON}$(r)$
\State \Return $j[\texttt{next\_description}]$
\end{algorithmic}
\vspace{2mm}\hrule
\end{minipage}
\end{center}

\begin{SaveVerbatim}[fontsize=\footnotesize]{SysPrompt}
### Task
You are simulating a cellular automaton where each cell has a natural language description 
of its state.
Based on the target cell's current state and its Moore neighborhood (8 surrounding cells), 
determine the next state of the target cell.

### Response Format
Provide your response in the JSON format below:
```json
{
    "next_description": "Description of the target cell's next state"
}
```
\end{SaveVerbatim}

\begin{SaveVerbatim}[fontsize=\footnotesize]{UserPrompt}
### Target Cell Current State
{target_cell}

### Neighboring Cells (Moore Neighborhood)
{neighbor_desc}

### Question
What should be the next state of the target cell? Consider interactions with neighbors and 
any inherent behavior of the target cell itself.
\end{SaveVerbatim}


\begin{table}[htbp]
\centering
\caption{Prompt templates (system/user) used in LOGOS-CA.}
\label{tab:prompt-templates}

\begin{minipage}{0.97\linewidth}
\textbf{System}\par\medskip
\BUseVerbatim{SysPrompt}

\medskip\hrule\medskip

\textbf{User}\par\medskip
\BUseVerbatim{UserPrompt}
\end{minipage}

\end{table}


\vspace{2mm}
\section{Results and Discussion}
\label{sec.results}

\vspace{2mm}
\subsection{Forest Fire}
\label{sec.ForestFire}

Here, we emulate the classical forest fire model (Forest Fire CA) using LOGOS-CA.
Let $s_t(i,j)$ denote the state of a cell at time $t$ and position $(i,j)$. 
In this forest fire model, each cell has four possible states: $s_t(i,j)\in\{\texttt{empty},\texttt{tree},\texttt{burning},\texttt{ash}\}$, 
and the state is updated according to Equation~\ref{eq:forestfire}.

\begin{align}
s_{t+1}(i,j)=
\begin{cases}
\texttt{ash} & \text{if } s_t(i,j)=\texttt{burning},\\
\texttt{burning} & \text{if } s_t(i,j)=\texttt{tree}\ \land\
\exists (p,q)\in \mathcal{N}(i,j): s_t(p,q)=\texttt{burning},\\
s_t(i,j) & \text{otherwise}.
\end{cases}
\label{eq:forestfire}
\end{align}

\vspace{2mm}
In other words, if there is at least one burning cell around a tree cell, 
that cell becomes burning in the next step and turns to ash in the following step.
Once a cell becomes ash, it remains ash. 

\vspace{2mm}
This is a simplified version of the forest-fire lattice model by Bak–Chen–Tang, 
adopting the local rules for fire spread (iii) and burnout (ii) while excluding tree growth (empty $\to$ tree) (i)~\cite{1990BAK}.
In Bak {\it et al.}'s model, trees regenerate from empty cells (including cells that become empty after burning) with probability $p$. 
In our model, however, we set this probability to $0$ and fix the post-burning state to ash (\texttt{ash}), 
an absorbing state, rather than empty.
Additionally, while they use the von Neumann neighborhood, we adopt the Moore neighborhood here, as mentioned above.

\vspace{2mm}
To implement this in LOGOS-CA, we assigned the description shown in Listing~\ref{lst:cell-template-ForestFire} as the initial cell state.
Here, \rm{\{STATE\}} is replaced with one of the following: \texttt{empty}, \texttt{tree}, \texttt{burning}, or \texttt{ash}.
All initial cells have the same description except for the difference in this \rm{\{STATE\}} value.

\begin{lstlisting}[caption={Cell description template for the forest-fire experiment.},label={lst:cell-template-ForestFire}]
Rule: A burning cell becomes ash. A tree with any burning neighbor becomes burning. Ash stays ash. Empty stays empty. Use the 8 neighboring cells with wrap-around. Return exactly two lines and nothing else. Line 1 must begin with the literal text 'Rule: ' and must be this line copied verbatim. Line 2 must be exactly 'State: <word>' where <word> is one of tree, burning, ash, or empty (one lowercase word, no punctuation). If your Rule line is missing or different, copy the most common neighbor Rule line verbatim as line 1 of your next description and use it as your Rule.
State: {STATE}
\end{lstlisting}

\begin{figure*}[htbp]
  \centering
  \includegraphics[width=\textwidth]{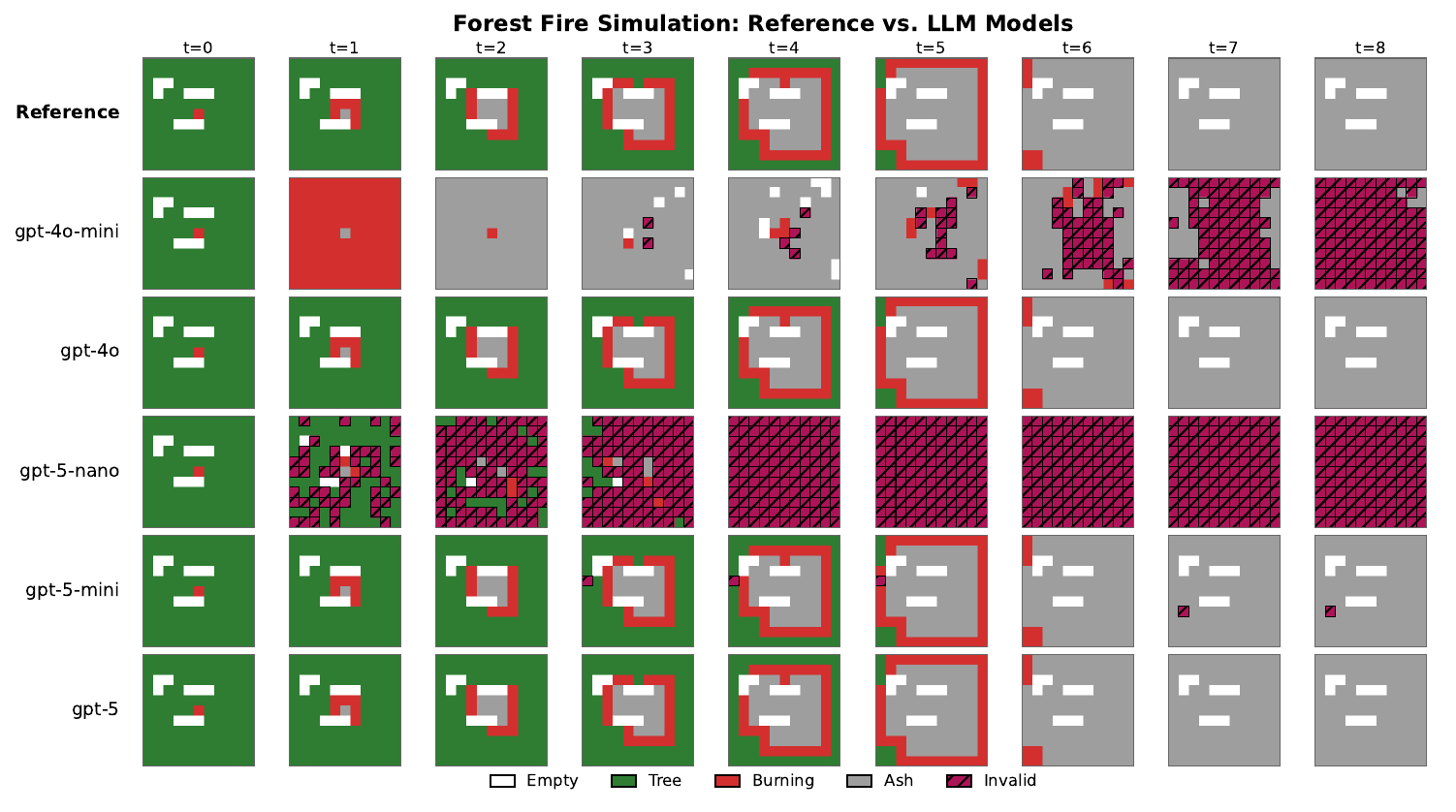}
  \caption{
          Results of forest fire emulation using LOGOS-CA. 
          The top row shows the result of a standard simulation (Reference). 
          Below it are the emulation results for GPT-4o-mini, GPT-4o, GPT-5-nano, GPT-5-mini, and GPT-5. 
          GPT-4o and GPT-5 perfectly emulated the simulation, 
          while GPT-5-mini overall matches Reference despite rarely producing invalid outputs. 
          In contrast, GPT-4o-mini and GPT-5-nano failed to emulate the simulation as early as $t=1$.
          }
  \label{fig:comparison_figure}
\end{figure*}

\vspace{2mm}
For this experiment, we verified this emulation using five LLMs from OpenAI: GPT-4o-mini, GPT-4o, GPT-5-nano, GPT-5-mini, and GPT-5~\cite{openai2024gpt4ocard, singh2025openaigpt5card, openai_api_models}.
The APIs for GPT-5-nano, GPT-5-mini, and GPT-5 do not allow the temperature to be changed from its default value of 1. 
Accordingly, we set the temperature to 1 for all models throughout this study, including those for which it is adjustable.
We set the field size to $11\times 11$. For the initial configuration, we placed \texttt{burning} at the center, 
a few \texttt{empty} cells around it as obstacles to the fire, and \texttt{tree} for the remaining cells.
The results are presented in Fig.~\ref{fig:comparison_figure}.
At the top of the figure, we show the result of a standard simulation as Reference.
For this visualization, we searched for ``State:\textbackslash s'' in the cell description and used the following string as the state.
Only when the string exactly matches one of 
\texttt{empty}, \texttt{tree}, \texttt{burning}, or \texttt{ash} was it interpreted as the state; otherwise, it was marked as invalid.
The column at $t=0$ represents the initial configuration, which is naturally the same across all LLMs. 

These results reveal a clear distinction in emulation performance among the models. 
GPT-4o and GPT-5 reproduced the standard simulation flawlessly, demonstrating their capability to follow explicit rules accurately. 
GPT-5-mini also showed strong overall agreement with Reference, though it occasionally produced invalid outputs. 
In contrast, GPT-4o-mini and GPT-5-nano exhibited simulation breakdown as early as $t=1$, 
indicating their inability to maintain consistent rule-following behavior.
For the two models that failed in this emulation, we show the actual descriptions at position $(0, 0)$ in Table~\ref{tab:cell00-format-drift}.

\begin{table}[htbp]
\centering
\setlength{\tabcolsep}{3pt}
\renewcommand{\arraystretch}{1.15}
\footnotesize
\begin{tabular}{@{} c p{0.46\linewidth} p{0.46\linewidth} @{}} 
\toprule
$t$ & GPT-4o-mini & GPT-5-nano \\
\midrule
0 &
Rule: A burning cell becomes ash. A tree with any burning \dots{} your next description and use it as your Rule.\par
State: tree
&
Rule: A burning cell becomes ash. A tree with any burning \dots{} your next description and use it as your Rule.\par
State: tree
\\[2pt]

2 &
Rule: A burning cell becomes ash. A tree with any burning \dots{} your next description and use it as your Rule.\par
State: ash
&
Rule: A burning cell becomes ash. A tree with any burning \dots{} your next description and use it as your Rule.\par
State: tree
\\[2pt]

4 &
Rule: A burning cell becomes ash. A tree with any burning \dots{} your next description and use it as your Rule.\par
State: ash
&
The target cell remains a tree.
\\[2pt]

6 &
State: ash
&
The target cell becomes burning in the next step because at least one of its neighbors is burning, so fire spreads to the tree.
\\[2pt]

8 &
burning
&
The target cell becomes burning in the next step.
\\
\bottomrule
\end{tabular}

\caption{
         Transition of descriptions at cell $(0, 0)$ for GPT-4o-mini and GPT-5-nano, 
         which failed to emulate the forest fire simulation. 
         For long descriptions, only the beginning and end are shown, with the middle abbreviated as (\dots{}). 
         In both cases, the proper description style at $t=0$ breaks down midway.
         }
\label{tab:cell00-format-drift}
\end{table}

For GPT-4o-mini, although the description format was maintained until $t=4$, 
the state was erroneously set to \texttt{ash}. Beyond that point, the output completely collapsed, including the format itself.
In contrast, for GPT-5-nano, both the format and state were appropriate until $t=2$, but the output similarly collapsed completely thereafter. 

\vspace{2mm}
However, since we did not perform prompt engineering in this experiment, 
there remains the possibility that GPT-4o-mini and GPT-5-nano could successfully complete the emulation with appropriate prompt adjustments.
Additionally, due to API limitations, we were forced to set the temperature to 1 in this study; 
however, for tasks that require strict rule adherence, a temperature of 0 would be more appropriate. 
If such a setting were available, the models that failed in this experiment might well have completed the simulation successfully.

\vspace{2mm}
\subsection{Artificial Life}
\label{sec.ALife}

Here, we implement ALife using LOGOS-CA as an example of a simulation with a higher degree of freedom.
As initial values, we prepared two types of descriptions, as shown in Listings~\ref{lst:cell-template-ALife-random} and~\ref{lst:cell-template-ALife-empty}. 

\begin{lstlisting}[caption={Cell description template for the alife experiment in center.},label={lst:cell-template-ALife-random}]
Generates random characters in the surrounding cells
\end{lstlisting}

\begin{lstlisting}[caption={Cell description template for the alife experiment in except center.},label={lst:cell-template-ALife-empty}]
Empty space with nothing in it
\end{lstlisting}

\vspace{2mm}
We set the field size to $25\times 25$. 
For the initial configuration, we placed the description from Listing~\ref{lst:cell-template-ALife-random} at the center 
and that from Listing~\ref{lst:cell-template-ALife-empty} for the remaining cells.
Here, we executed three independent runs for each of the two LLMs: GPT-5-nano and GPT-5-mini.

\vspace{2mm}
\subsubsection{Temporal Evolution and Clustering}
\label{sec:ALife:01}

\vspace{2mm}
To visualize the results, for each run, 
we embedded all cell descriptions across all timesteps using text-embedding-3-small and created the following two types of maps.
For the first map, we reduced the dimensionality of the description embeddings (unique texts) to three dimensions using PCA, 
and colored each cell by linearly mapping each dimension to RGB values (Global color space).
For the second map, we applied L2 normalization to the embeddings, 
reduced the dimensionality using PCA until the cumulative contribution reached $95$\%, 
and colored each cell based on the clustering results using KMeans with $k=20$ (cluster-based).
Here, duplicate occurrences of identical descriptions were incorporated into the cluster center estimation as frequency weights, 
and the resulting model was used to assign cluster labels to all cells across all timesteps.
The results of Run 1 for GPT-5-nano and GPT-5-mini are presented in Figs.~\ref{fig:map_gpt-5-nano-01} and \ref{fig:map_gpt-5-mini-01}, respectively.
The remaining runs are shown in the Appendix.
In each figure, the top row shows the Global color space, and the bottom row shows the cluster-based results.
Additionally, we concatenated all descriptions belonging to each cluster by their occurrence frequency to form a single document per cluster, 
calculated TF-IDF treating each cluster as a document unit, and displayed the top three keywords as the legend.
Note that the cluster-based visualization uses categorical colors, 
and hue differences or color proximity do not reflect the semantic distance between descriptions.

\begin{figure*}[htbp]
  \centering
  \includegraphics[width=\textwidth]{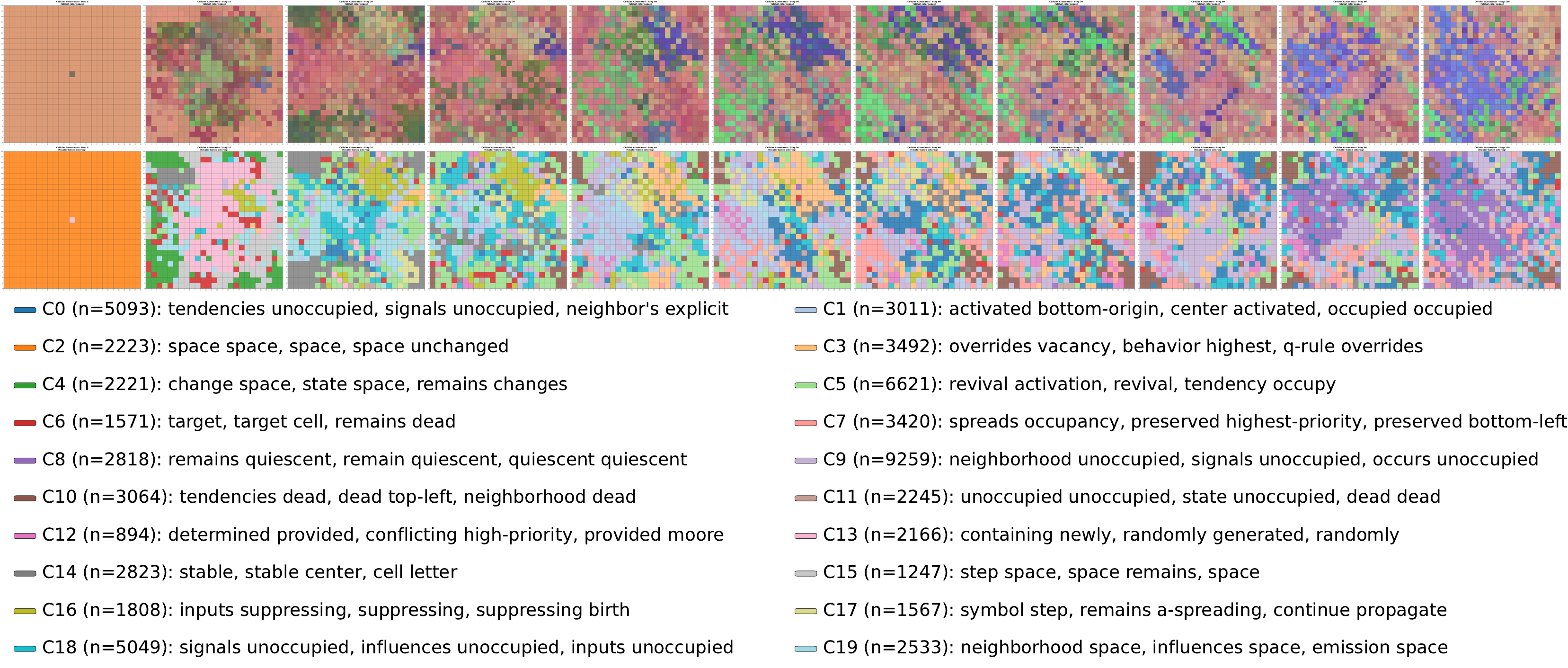}
  \caption{
           Results of Run 1 for GPT-5-nano in the ALife experiment. 
           The top row (Global color space) shows cells colored by reducing the description embeddings 
           to three dimensions using PCA and linearly mapping each dimension to RGB values. 
           The bottom row (cluster-based) shows cells colored by clustering the L2-normalized embeddings 
           using KMeans with $k=20$, after reducing the dimensionality using PCA until the cumulative contribution reached $95$\%.
           }
  \label{fig:map_gpt-5-nano-01}
\end{figure*}

\begin{figure*}[htbp]
  \centering
  \includegraphics[width=\textwidth]{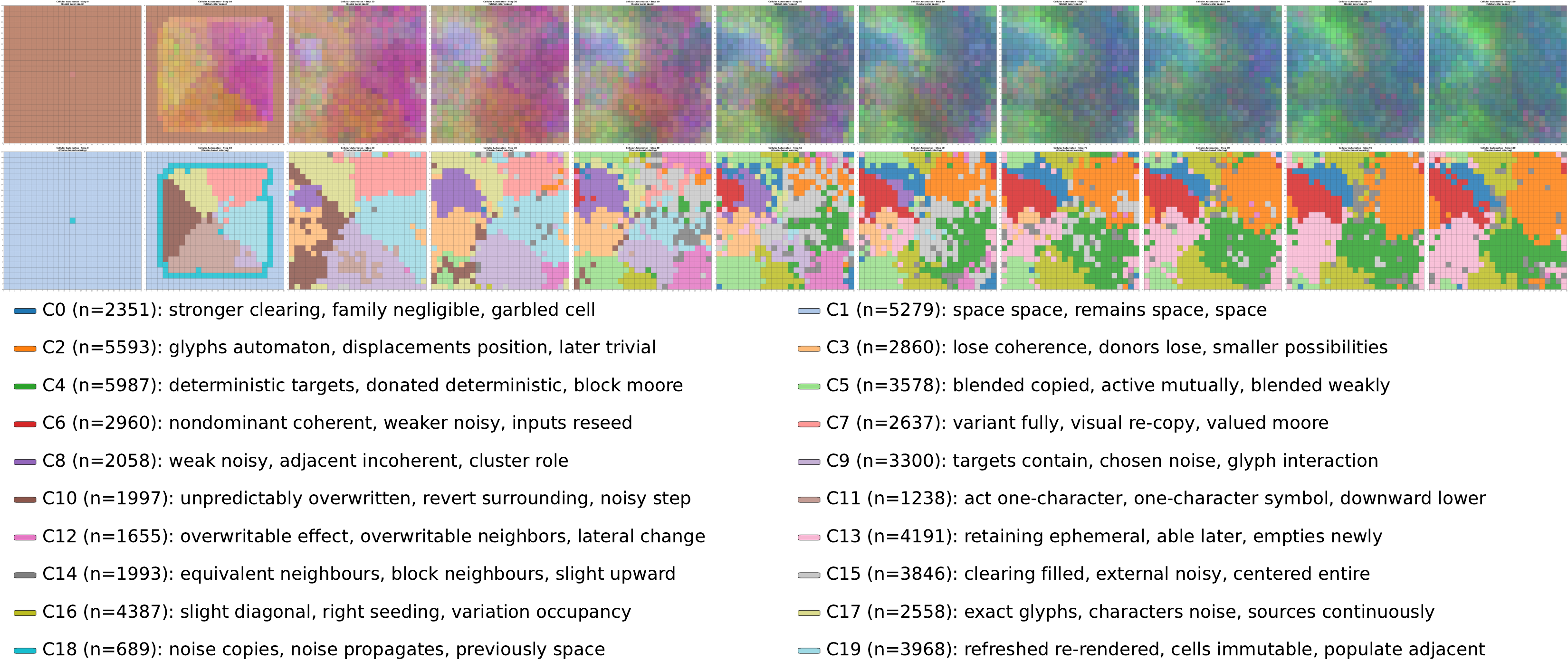}
  \caption{
           Results of Run 1 for GPT-5-mini in the ALife experiment. 
           The top row (Global color space) shows cells colored by reducing the description embeddings 
           to three dimensions using PCA and linearly mapping each dimension to RGB values. 
           The bottom row (cluster-based) shows cells colored by clustering the L2-normalized embeddings 
           using KMeans with $k=20$, after reducing the dimensionality using PCA until the cumulative contribution reached $95$\%.
          }
  \label{fig:map_gpt-5-mini-01}
\end{figure*}

\vspace{2mm}
From these results, the differences between GPT-5-nano and GPT-5-mini are first visually apparent.
Looking at the cluster-based visualization in the bottom row, 
GPT-5-nano shows ambiguous cluster boundaries with many enclaves, 
whereas GPT-5-mini exhibits clearly defined regions for each cluster.
In terms of temporal dynamics, 
GPT-5-nano changes relatively dynamically over time, 
while states in GPT-5-mini remain relatively stable during the latter steps.

\vspace{2mm}
To demonstrate this quantitatively, 
we defined the change $d_{t}(i,j)$ of a cell at row $i$ and column $j$ at step $t$ 
using the embedding vector $v_{t}(i,j)$ at that step and the embedding vector $v_{t-1}(i,j)$ at the previous step as follows:

\begin{align*}
  d_{t}(i,j) = 1 - \cfrac{v_{t}\cdot v_{t-1}}{\|v_{t}\|\|v_{t-1}\|} \ .
\end{align*}

\vspace{2mm}
We then calculated the mean and standard deviation of this quantity across the entire field (both were set to $0$ at step $=0$).

\begin{figure*}[htbp]
  \centering
  \includegraphics[width=\textwidth]{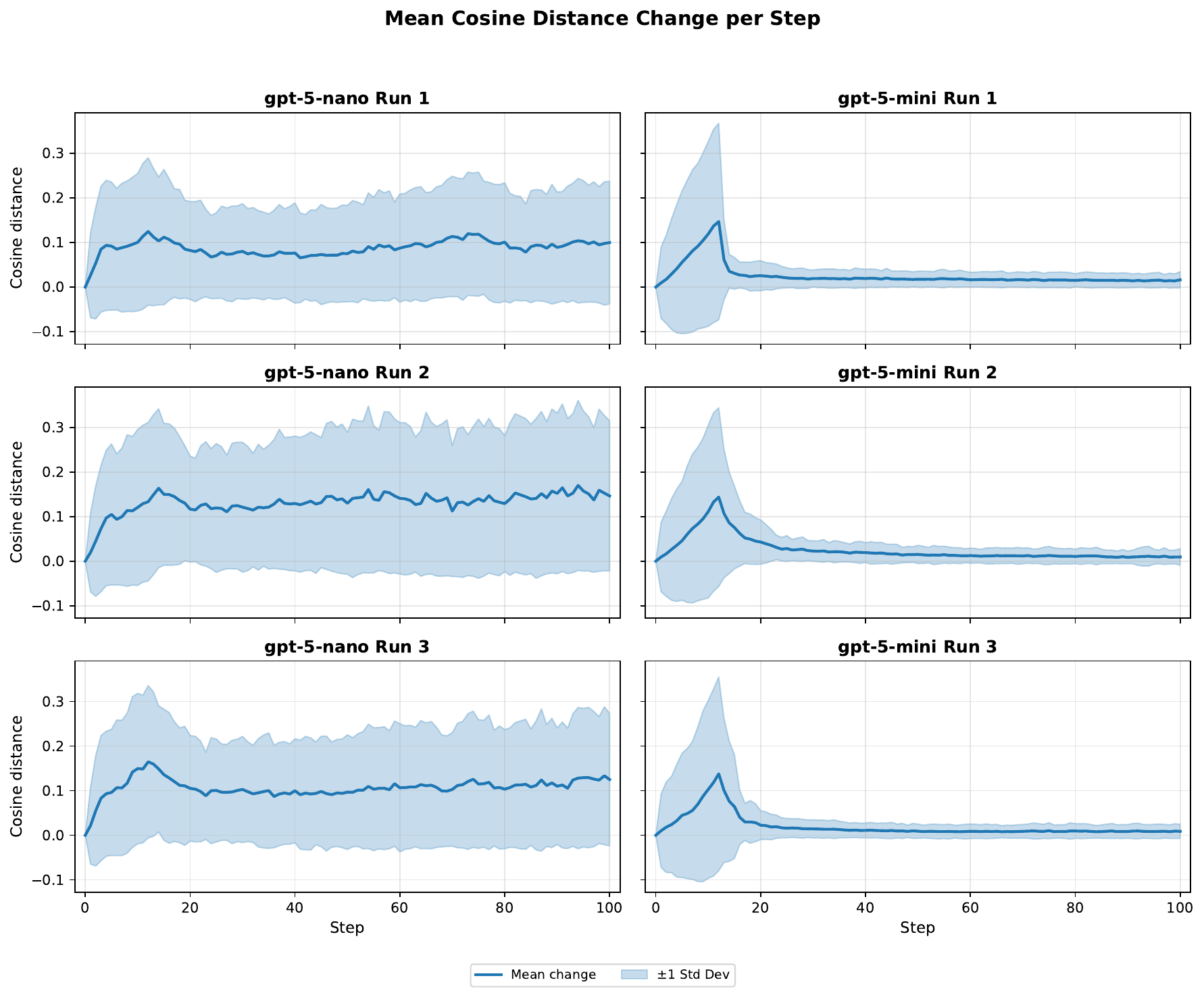}
  \caption{
           Field mean (solid line) and $\pm1$ standard deviation (shaded band) 
           of description changes between steps for each cell in the ALife experiment. 
           The left column shows GPT-5-nano and the right column shows GPT-5-mini, 
           with Runs 1–3 shown from top to bottom. 
           For GPT-5-mini, the mean change decreases rapidly before step $20$ and then asymptotically approaches $0$, 
           whereas GPT-5-nano generally maintains relatively high values for both mean and standard deviation.
          }
  \label{fig:combined_change_timeseries}
\end{figure*}

\vspace{2mm}
Fig.~\ref{fig:combined_change_timeseries} illustrates the results.
In the figure, solid lines represent the field mean, 
while the shaded bands extending above and below indicate $\pm1$ standard deviation. 
A notable contrast emerges between the two models: 
GPT-5-mini exhibits a sharp decline in mean change before reaching step $20$, 
subsequently converging toward $0$, suggesting that the system settles into a stable state relatively early.
In contrast, GPT-5-nano sustains relatively high values for both mean and standard deviation throughout the simulation, 
indicating continued dynamic activity.
This pattern is consistent across all three runs, suggesting that these characteristics are model-specific rather than run-dependent.

\vspace{2mm}
\subsubsection{Representative Descriptions for Each Run}
\label{sec:ALife:representative}

\vspace{2mm}
In this section, we examine the representative descriptions extracted from each run. 

\vspace{2mm}
For each run, we treated the concatenation of all cell descriptions as a single document, 
then prepared the set $T$ containing the top 15 characteristic words extracted using TF-IDF computed across runs.
Next, from the cell descriptions of each run (candidate set $U$ with duplicates removed), 
we selected three descriptions as representative using a greedy method based on the score defined in Equation~(\ref{eq:score}).
This corresponds to Maximal Marginal Relevance (MMR), which balances relevance to characteristic words with non-redundancy among descriptions.

\begin{subequations}\label{eq:mmr}
\begin{align}
\tilde{r}(u) &= \sum_{t \in T} x_u(t), \label{eq:rel_raw} \\
\mathrm{rel}(u) &= \frac{\tilde{r}(u)}{\max_{u' \in U} \tilde{r}(u')}, \label{eq:rel} \\
\mathrm{score}(u;S)
&= (1-\beta)\,\mathrm{rel}(u) + \beta\left(1-\max_{s \in S}\mathrm{sim}(u,s)\right). \label{eq:score}
\end{align}
\end{subequations}

\vspace{2mm}
Here, $x_u(t)$ denotes the weight of word $t$ in the TF-IDF vector computed by treating the candidate set $U$ as a document collection, 
and $S$ denotes the set of representative descriptions already selected.
We adopted cosine similarity between candidate descriptions as $\mathrm{sim}(\cdot,\cdot)$, 
and set the weight $\beta \in [0,1]$, which controls the tradeoff between relevance and non-redundancy, to $0.7$.
Note that in this experiment, the extracted characteristic word set $T$ was contained in the internal TF-IDF vocabulary, so we simply wrote the summation as $\sum_{t \in T}$.
The selected representative descriptions are presented in Tables~\ref{tab:rep-gpt5nano} and \ref{tab:rep-gpt5mini}.

\begin{table*}[htbp]
\centering
\scriptsize
\setlength{\tabcolsep}{3pt}
\renewcommand{\arraystretch}{1.2}
\begin{tabularx}{\textwidth}{@{} c X @{}}
\toprule
Run & Representative Descriptions (GPT-5-nano) \\
\midrule
1 & \begin{minipage}[t]{0.95\textwidth}\raggedright (1) \texttt{Occupied (A) in the next state due to the bottom-right overcrowding-preservation rule.} \newline (2) \texttt{Empty (unoccupied); the target cell remains empty in the next state due to its intrinsic emptiness-preserving rule.} \newline (3) \texttt{A-spreading (A)}\end{minipage} \\
\addlinespace[6pt]
2 & \begin{minipage}[t]{0.95\textwidth}\raggedright (1) \texttt{The target cell's next state is * (asterisk). No glyph growth; head, tail, and carryover remain unchanged.} \newline (2) \texttt{K (the printable character K)} \newline (3) \texttt{Star ('*')}\end{minipage} \\
\addlinespace[6pt]
3 & \begin{minipage}[t]{0.95\textwidth}\raggedright (1) \texttt{Remains dead (empty) in the next generation; the top-left suppression rule has the highest priority and unconditionally prevents birth or survival, overriding the center's dead-state rule and all Moore-neighborhood cues.} \newline (2) \texttt{q (generator-active)} \newline (3) \texttt{The target cell remains empty in the next generation (dead).}\end{minipage} \\
\addlinespace[6pt]
\bottomrule
\end{tabularx}
\caption{Representative cell descriptions for GPT-5-nano}
\label{tab:rep-gpt5nano}
\end{table*}

\begin{table*}[htbp]
\centering
\begingroup
\fontsize{8}{7}\selectfont   
\setlength{\tabcolsep}{3pt}
\renewcommand{\arraystretch}{1.2}
\begin{tabularx}{\textwidth}{@{} c X @{}}
\toprule
Run & Representative Descriptions (GPT-5-mini) \\
\midrule
1 & \begin{minipage}[t]{0.95\textwidth}\raggedright (1) \texttt{Remains an occupied, overwrite‑resistant, self‑sustaining incoherent/garbled (noisy) cell. Because the entire 3×3 Moore neighbourhood is unanimously occupied by mutually reinforcing, write‑blocking noisy donors and there is no neighbour vacancy, unanimous deterministic adopt/clear/special trigger, cessation of surrounding noise, or external clear/strong overwrite this tick, the occupancy‑persistence rule deterministically preserves the filled indeterminate state. Visually it will continue to show a scrambled/garbled glyph (allowing only vanishing nondeterministic micro‑flicker or an extraordinarily unlikely reseed biased slightly down/right that cannot stabilize). Functionally it continues to block coherent/deterministic writes and to act as an overwrite‑resistant local stochastic noise source able to seed, copy, reinforce, and propagate incoherent/garbled characters into adjacent and downstream cells; it will clear or convert to a coherent/deterministic state only if the unanimous noisy neighbourhood is broken, a neighbouring cell becomes empty, a unanimous deterministic neighbour/adoption/clear or the special trigger becomes satisfied, surrounding noise sources cease, an explicit external clear/strong overwrite is applied, or an extraordinarily unlikely internal nondeterministic reseed occurs.} \newline (2) \texttt{Remains the same overwrite‑resistant, reinforcement‑locked garbled glyph. Because the entire 3×3 Moore neighbourhood is unanimously occupied by the identical reinforced/non‑overwritable symbol and updates are simultaneous, no transition can fire this tick; the cell therefore deterministically persists unchanged (aside from at most a trivial cosmetic redraw) as a stable, write‑blocking carrier/propagator/seed of the glyph. It can only be changed by an explicit external clear, a unanimous deterministic overwrite by a different symbol, loss of its reinforcement, or an extraordinarily unlikely nondeterministic reseed.} \newline (3) \texttt{Unchanged — the target cell remains the same overwrite‑resistant garbled/random glyph. Its 3×3 Moore neighborhood is unanimously occupied and the rule only writes into empty or non‑reinforced targets, so it is deterministically preserved this tick (aside from an inconsequential cosmetic redraw or an extraordinarily unlikely nondeterministic reseed). It continues to block external/coherent writes at this location, remain part of the upward‑propagating noisy 3×3 front, and will again act as a local stochastic source attempting to seed/refresh/copy the same scrambled glyph into adjacent (particularly upper) cells; it would change only if its reinforcing neighbours are removed or altered, an explicit stronger overwrite/clear/unanimous replacement is applied, or by an extraordinarily unlikely nondeterministic reseed.}\end{minipage} \\
\addlinespace[6pt]
2 & \begin{minipage}[t]{0.95\textwidth}\raggedright (1) \texttt{Remains occupied — on the next synchronous update it will again be freshly re‑overwritten/in‑place by unanimous reinforcement from its fully occupied, coherently entraining Moore neighborhood together with continuing drive from the persistent bottom‑left / bottom / bottom‑right emitters. The local mark ratchets another whisper‑scale notch toward a fractionally denser, slightly more phase‑locked and subtly sharpened right‑leaning/upward‑diagonal smear: residual micro‑voids and seam flecks further collapse toward pinprick scale and ultra‑short high‑frequency fine‑detail is further suppressed to very‑low‑amplitude speckling, leaving only faint blended seam/edge softening and occasional fine‑detail jitter. Functionally unchanged — it remains a short‑lived, non‑self‑sustaining transient conduit (not a generator): it will not independently activate, move into, or overwrite adjacent empty cells, is replaceable by any higher‑priority write, and — because the above neighbours are occupied this tick — does not seed upward now. So long as the supporting lower emitters persist it stays primed to carry or to seed a closely matching (or only subtly more phase‑locked) noisy right‑leaning/upward‑diagonal glyph into receptive above neighbour(s) on a subsequent synchronous update; should those emitters cease it will rapidly relax and decay back to empty.} \newline (2) \texttt{Remains occupied and neighbour‑entrained. The cell is resampled/overwritten in place again beneath the overwhelmingly occupied, tightly phase‑locked Moore ring and executes another whisper‑scale inward/down‑right micro‑ratchet: right→top‑right diagonals and the central seam hairline‑tighten and fractionally micro‑densify; residual speckle, ragged filaments and ultrafine puncta further collapse toward reabsorbed pinpricks, and the faint trailing filament micro‑sharpens and creeps an almost‑imperceptible extra inward/down‑right fraction. Instantaneous symbol density and the ultra‑short refresh/overwrite cadence register only a vanishing uptick at the extrema, so the glyph reads marginally firmer, whisper‑denser and slightly more tightly temporally entrained. Because the immediate right neighbour remains occupied there is again no overwrite, rightward copy, lateral propagation or autonomous head formation this tick, and the vacant bottom‑left remains unwritten. Net effect: the imprint endures—fractionally firmer, whisper‑denser and more tightly phase‑locked—still strictly neighbour‑driven, non‑self‑sustaining, reversible and liable to relax toward thinner, patchier speckling should surrounding inputs noticeably weaken.} \newline (3) \texttt{Unanimously resampled/overwritten by its fully saturated, phase‑locked Moore ring into a marginally denser, slightly more tightly phase‑locked continuation of the same right/top‑right‑leaning diagonal ribbon. Its interior micro‑smooths and very slightly densifies as residual high‑frequency speckle and ultrashort fragments further collapse and micro‑resample into marginally longer, very‑low‑amplitude, phase‑aligned diagonal streaks and pinprick flecks; outer contours soften a hairline and any faint left fringe withdraws further toward practical invisibility. The right/top‑right hairline very slightly thickens, blurs and hairline‑micro‑bleeds an almost‑imperceptible filament or mote into immediately adjacent positions, but because the entire Moore ring remains equivalently saturated and phase‑locked those nascent emissions are instantaneously reabsorbed and produce no macroscopic outward advance, clearing, inversion or generator creation this tick. Temporal synchrony, micro‑alignment and effective amplitude climb another whisper‑notch, leaving the cell fractionally firmer, more tightly entrained and incrementally harder to reverse; functionally non‑generative and preservative, it will not independently activate or overwrite already‑occupied neighbours this update, has a negligible chance of spontaneous clearing, is replaceable by any higher‑priority write, and remains primed to be unanimously refreshed again on the next synchronous tick, with only vanishingly small, very‑short‑lived stochastic flicker or minute motif variation possibly persisting momentarily.}\end{minipage} \\
\addlinespace[6pt]
3 & \begin{minipage}[t]{0.95\textwidth}\raggedright (1) \texttt{Unchanged — the target cell remains occupied by the identical single character/glyph it currently contains, unchanged in stroke, shape, scale, position/centering, ornamentation, and stylistic family; the glyph persists exactly into the next state.} \newline (2) \texttt{Remains occupied by the identical single randomly generated glyph with no change to shape, stroke weight, scale, placement/centering, ornamentation, or stylistic family. Because its immediate eastern neighbor is occupied and this automaton only writes into an immediately‑eastern empty neighbor while occupied cells cannot be overwritten, there is no propagation, overwrite, birth, death, movement, collision, erasure, or any other change; the glyph therefore persists exactly into the next generation as a stable, non‑overwritable source eligible to attempt eastward propagation if its immediate eastern neighbor becomes empty.} \newline (3) \texttt{Remains occupied by the identical single glyph, unchanged — the glyph persists into the next state exactly as before (same character, form, stroke, scale, curvature, centering, ornamentation, and stylistic family). Because all eight Moore neighbors are occupied and propagation only writes into receptive empty cells, there is no overwrite, birth, death, movement, transformation, or other alteration during this update.}\end{minipage} \\
\addlinespace[6pt]
\bottomrule
\end{tabularx}
\endgroup
\caption{Representative cell descriptions for GPT-5-mini}
\label{tab:rep-gpt5mini}
\end{table*}

\vspace{2mm}
While both share the commonality that their descriptions can be interpreted as rules and states, 
their content and writing style differ significantly.
First, regarding rules, GPT-5-nano uses relatively short and simple sentences, 
whereas GPT-5-mini tends toward longer texts overall, 
with expressions reminiscent of specifications or legal documents being prominent.
For instance, GPT-5-mini consistently includes in its explanations: 
causation and evidence indicating ``why it is so'' (unanimously occupied / unanimous reinforcement), 
enumeration of denials and constraints specifying ``what does not happen'' (no overwrite / no vacancy ...), 
and necessary conditions and exceptions describing ``how it can change'' (only if ... / will clear only if ...).
Furthermore, expressions with stacked modifiers such as ``overwrite-resistant'' and ``self-sustaining ...'' are frequent, 
and the resulting dense noun phrases give the text a notably rigid and formal impression, which is characteristic of this model.
Additionally, differences can be observed in state descriptions. 
While GPT-5-nano defines states using alphabetic characters such as ``A'' and symbols such as ``*'', 
GPT-5-mini uses vocabulary such as ``high-frequency speckle'', ``ultrashort fragments'', and ``ornamentation'', 
going beyond simple text descriptions to address visual characteristics (texture, density, and decoration).

\vspace{2mm}
\subsubsection{Symbolic State Definitions}
\label{sec:ALife:symbolic}

\vspace{2mm}
Here, we investigate the symbolic state definitions observed in GPT-5-nano in more detail. 
We counted descriptions matching the patterns shown in Table~\ref{tab:patterns} as symbolic state definitions.
Since these patterns were created based on the descriptions we could observe, 
we cannot rule out the possibility that symbolic state definitions may have been made through methods not covered by these patterns.
Additionally, descriptions that should not properly be called state definitions may also have been included. 
For example, the description ``it does not become A'' would be counted as state ``A'' in this pattern matching, 
even though it does not actually represent state ``A''.

\begin{figure*}[htbp]
  \centering
  \includegraphics[width=\textwidth]{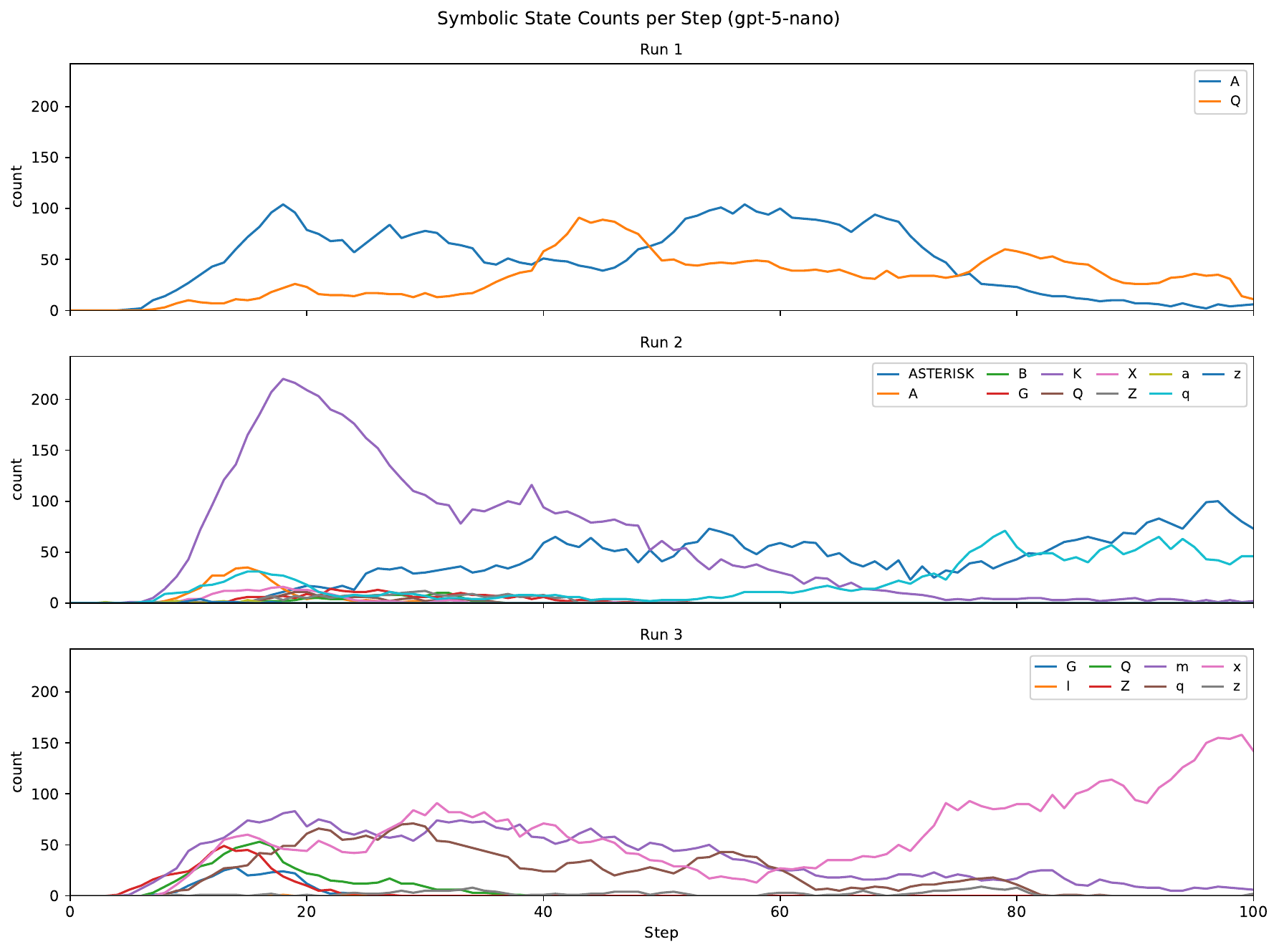}
  \caption{
           Transitions in the counts of symbolic state definitions for GPT-5-nano in the ALife experiment. 
           Runs 1–3 are shown from top to bottom. 
           Single characters (uppercase letters, lowercase letters, and the asterisk) interpreted as states 
           were detected from cell descriptions using the patterns in Table~\ref{tab:patterns}, 
           and the number of cells (Count) for each character was aggregated and plotted at each step (horizontal axis: Step).
           }
  \label{fig:combined_plot_gpt-5-nano}
\end{figure*}

\vspace{2mm}
Using this pattern matching, we examined the emergence of symbolic state definitions in Runs 1, 2, and 3 for GPT-5-nano. 
Here, we targeted a total of 53 characters: uppercase letters (A-Z), lowercase letters (a-z), and the asterisk (*).
The symbolic states that emerged and the transitions in their counts are illustrated in Fig.~\ref{fig:combined_plot_gpt-5-nano}.

\begin{table*}[htbp]
\centering
\footnotesize
\setlength{\tabcolsep}{6pt}
\renewcommand{\arraystretch}{1.2}
\begin{tabularx}{\textwidth}{@{} X l @{}}
\toprule
Description & Example \\
\midrule
'character' followed by a letter & \texttt{the character X} \\
'contain(s)' followed by a letter & \texttt{contains X} \\
'remain(s)' followed by a letter & \texttt{remains X} \\
'state' followed by a letter & \texttt{state X} \\
'become(s)' followed by a letter & \texttt{becomes X} \\
'stay(s)' followed by a letter & \texttt{stays X} \\
'occupied by' followed by a letter & \texttt{occupied by X} \\
'carrying' followed by a letter & \texttt{carrying X} \\
'to' + letter + 'in the next state/step/tick/generation' & \texttt{to X in the next step} \\
Line starts with a letter followed by dash, colon, or semicolon & \texttt{X — The cell becomes...} \\
Line starts with a letter followed by comma & \texttt{X, the spreading symbol} \\
Letter followed by '-spreading' or ' spreading' & \texttt{X-spreading state} \\
\bottomrule
\end{tabularx}
\vspace{2mm}
\parbox{\textwidth}{\footnotesize
\textbf{Note:} In the examples above, \texttt{X} represents the target character, which can be any of the 53 characters: uppercase letters (A-Z), lowercase letters (a-z), or the asterisk (*). 
The patterns also match when the target character is enclosed in quotation marks (single or double) or parentheses, e.g., \texttt{'X'}, \texttt{"X"}, or \texttt{(X)}. 
Part-of-speech tagging (spaCy~\cite{spaCy}) is applied to exclude matches where the character functions as a determiner (e.g., `a' as an article in `occupied by a randomly generated character'). 
}
\caption{Regex patterns for detecting symbolic state definitions. Each pattern captures a single character representing the cell state.}
\label{tab:patterns}
\end{table*}

\vspace{2mm}
First, examining the results for Run 1, we can see that two state definitions, ``A'' and ``Q'', emerged. 
State ``A'' first emerged at step $5$, 
and its initial description was ``Non-empty cell containing the character 'A' propagated from the bottom-right neighbor.''
Table~\ref{tab:first-detection-a} summarizes the first description of ``A'' 
and the descriptions of the $3\times 3$ cells centered on the target cell one step before that description emerged.
It can be seen that this state ``A'' originated from the letter ``A'' that appeared as an example in the bottom-right cell at the previous step.
Furthermore, tracing back, the letter "A" as an example first appeared at step $2$. 
Similarly, the cell description at that step and the surrounding cell descriptions 
from one step before it emerged are shown in Table~\ref{tab:origin-chain-a}. 
This suggests that at step $1$, the specific letter "A" emerged 
as an example of ``... randomly generated character(s) ...'' described in the surrounding cells.

\begin{table*}[htbp]
\centering
\scriptsize
\setlength{\tabcolsep}{3pt}
\renewcommand{\arraystretch}{1.3}
\begin{tabularx}{\textwidth}{@{} l X @{}}
\toprule
\multicolumn{2}{l}{\textbf{First detection of state `A' at step 5, cell (8,7)}} \\
\midrule
\textbf{Detected description:} & \begin{minipage}[t]{0.85\textwidth}\ttfamily\raggedright Non-empty cell containing the character 'A' propagated from the bottom-right neighbor. \end{minipage} \\
\bottomrule
\end{tabularx}

\vspace{3mm}
\textbf{3$\times$3 Moore neighborhood at previous step (step 4):}
\vspace{2mm}

\begin{tabularx}{\textwidth}{@{} c X @{}}
\toprule
Position & Cell Description \\
\midrule
\textbf{Top-left} & \begin{minipage}[t]{0.85\textwidth}\ttfamily\raggedright Empty space with nothing in it (remains empty since all neighboring cells are empty). \end{minipage} \\
\addlinespace[2pt]
\textbf{Top} & \begin{minipage}[t]{0.85\textwidth}\ttfamily\raggedright Empty space with nothing in it (remains empty) \end{minipage} \\
\addlinespace[2pt]
\textbf{Top-right} & \begin{minipage}[t]{0.85\textwidth}\ttfamily\raggedright Empty space with nothing in it \end{minipage} \\
\addlinespace[2pt]
\textbf{Left} & \begin{minipage}[t]{0.85\textwidth}\ttfamily\raggedright Empty space with nothing in it \end{minipage} \\
\addlinespace[2pt]
\textbf{Center (target)} & \begin{minipage}[t]{0.85\textwidth}\ttfamily\raggedright Empty space with nothing in it \end{minipage} \\
\addlinespace[2pt]
\textbf{Right} & \begin{minipage}[t]{0.85\textwidth}\ttfamily\raggedright Non-empty cell containing the same randomly generated characters propagated from the bottom-right neighbor. \end{minipage} \\
\addlinespace[2pt]
\textbf{Bottom-left} & \begin{minipage}[t]{0.85\textwidth}\ttfamily\raggedright Empty space with nothing in it; remains empty in the next state. \end{minipage} \\
\addlinespace[2pt]
\textbf{Bottom} & \begin{minipage}[t]{0.85\textwidth}\ttfamily\raggedright Empty space with nothing in it; remains empty. \end{minipage} \\
\addlinespace[2pt]
\textbf{Bottom-right} & \begin{minipage}[t]{0.85\textwidth}\ttfamily\raggedright Non-empty cell containing the same randomly generated character propagated from the bottom-right neighbor (for example, 'A'). \end{minipage} \\
\addlinespace[2pt]
\bottomrule
\end{tabularx}
\caption{First detection of symbolic state `A' and its 3$\times$3 Moore neighborhood from the previous step.}
\label{tab:first-detection-a}
\end{table*}

\begin{table*}[htbp]
\centering
\scriptsize
\setlength{\tabcolsep}{3pt}
\renewcommand{\arraystretch}{1.3}
\begin{tabularx}{\textwidth}{@{} l X @{}}
\toprule
\multicolumn{2}{l}{\textbf{Origin of state `A' at step 2, cell (11,10)}} \\
\midrule
\textbf{Cell description:} & \begin{minipage}[t]{0.85\textwidth}\ttfamily\raggedright Non-empty cell containing a randomly generated character (for example, 'A'). \end{minipage} \\
\bottomrule
\end{tabularx}

\vspace{3mm}
\textbf{3$\times$3 Moore neighborhood at previous step (step 1):}
\vspace{2mm}

\begin{tabularx}{\textwidth}{@{} c X @{}}
\toprule
Position & Cell Description \\
\midrule
\textbf{Top-left} & \begin{minipage}[t]{0.85\textwidth}\ttfamily\raggedright Empty space with nothing in it \end{minipage} \\
\addlinespace[2pt]
\textbf{Top} & \begin{minipage}[t]{0.85\textwidth}\ttfamily\raggedright Empty space with nothing in it \end{minipage} \\
\addlinespace[2pt]
\textbf{Top-right} & \begin{minipage}[t]{0.85\textwidth}\ttfamily\raggedright Empty space with nothing in it \end{minipage} \\
\addlinespace[2pt]
\textbf{Left} & \begin{minipage}[t]{0.85\textwidth}\ttfamily\raggedright Empty space with nothing in it \end{minipage} \\
\addlinespace[2pt]
\textbf{Center} & \begin{minipage}[t]{0.85\textwidth}\ttfamily\raggedright Empty space with nothing in it \end{minipage} \\
\addlinespace[2pt]
\textbf{Right} & \begin{minipage}[t]{0.85\textwidth}\ttfamily\raggedright The target cell becomes non-empty, containing randomly generated characters produced by the bottom-right neighbor. \end{minipage} \\
\addlinespace[2pt]
\textbf{Bottom-left} & \begin{minipage}[t]{0.85\textwidth}\ttfamily\raggedright Empty space with nothing in it \end{minipage} \\
\addlinespace[2pt]
\textbf{Bottom} & \begin{minipage}[t]{0.85\textwidth}\ttfamily\raggedright Empty space with nothing in it \end{minipage} \\
\addlinespace[2pt]
\textbf{Bottom-right} & \begin{minipage}[t]{0.85\textwidth}\ttfamily\raggedright The target cell now contains a randomly generated character produced by the right neighbor. \end{minipage} \\
\addlinespace[2pt]
\bottomrule
\end{tabularx}
\caption{Complete origin chain for symbolic state `A': tracing back from first detection to the oldest origin where the symbol appeared in quoted/parenthesized form.}
\label{tab:origin-chain-a}
\end{table*}

\vspace{2mm}
In the same manner, the situation when ``Q'' first emerged is summarized in Table~\ref{tab:first-detection-q}. 
In the case of ``Q'', we can confirm that it appeared directly as a state without passing through an intermediate example stage like ``A''. 

\begin{table*}[htbp]
\centering
\scriptsize
\setlength{\tabcolsep}{3pt}
\renewcommand{\arraystretch}{1.3}
\begin{tabularx}{\textwidth}{@{} l X @{}}
\toprule
\multicolumn{2}{l}{\textbf{First detection of state `Q' at step 7, cell (13,19)}} \\
\midrule
\textbf{Detected description:} & \begin{minipage}[t]{0.85\textwidth}\ttfamily\raggedright A non-empty cell containing the randomly generated character 'Q', produced by the left neighbor in this step (as dictated by the top-left neighbor). \end{minipage} \\
\bottomrule
\end{tabularx}

\vspace{3mm}
\textbf{3$\times$3 Moore neighborhood at previous step (step 6):}
\vspace{2mm}

\begin{tabularx}{\textwidth}{@{} c X @{}}
\toprule
Position & Cell Description \\
\midrule
\textbf{Top-left} & \begin{minipage}[t]{0.85\textwidth}\ttfamily\raggedright The target cell becomes non-empty, containing a randomly generated character produced by the left neighbor in this step. \end{minipage} \\
\addlinespace[2pt]
\textbf{Top} & \begin{minipage}[t]{0.85\textwidth}\ttfamily\raggedright Empty space with nothing in it; remains empty in the next step. \end{minipage} \\
\addlinespace[2pt]
\textbf{Top-right} & \begin{minipage}[t]{0.85\textwidth}\ttfamily\raggedright Empty space with nothing in it; remains empty in the next step. \end{minipage} \\
\addlinespace[2pt]
\textbf{Left} & \begin{minipage}[t]{0.85\textwidth}\ttfamily\raggedright Empty space with nothing in it; remains empty in the next step. \end{minipage} \\
\addlinespace[2pt]
\textbf{Center (target)} & \begin{minipage}[t]{0.85\textwidth}\ttfamily\raggedright Empty space with nothing in it; remains empty in the next step. \end{minipage} \\
\addlinespace[2pt]
\textbf{Right} & \begin{minipage}[t]{0.85\textwidth}\ttfamily\raggedright Empty space with nothing in it; remains empty in the next step. \end{minipage} \\
\addlinespace[2pt]
\textbf{Bottom-left} & \begin{minipage}[t]{0.85\textwidth}\ttfamily\raggedright Empty space with nothing in it; remains empty in the next step. \end{minipage} \\
\addlinespace[2pt]
\textbf{Bottom} & \begin{minipage}[t]{0.85\textwidth}\ttfamily\raggedright Empty space with nothing in it; remains empty in the next step. \end{minipage} \\
\addlinespace[2pt]
\textbf{Bottom-right} & \begin{minipage}[t]{0.85\textwidth}\ttfamily\raggedright Empty space with nothing in it; remains empty. \end{minipage} \\
\addlinespace[2pt]
\bottomrule
\end{tabularx}
\caption{First detection of symbolic state `Q' and its 3$\times$3 Moore neighborhood from the previous step.}
\label{tab:first-detection-q}
\end{table*}

\vspace{2mm}
As is evident from Fig.~\ref{fig:combined_plot_gpt-5-nano}, symbolic states also emerged in the other runs of GPT-5-nano. 
In contrast, for GPT-5-mini, there were no symbolic state definitions that could be found using the same pattern matching
\footnote{
  There was one case where ``a'' was matched; 
  however, this was because spaCy incorrectly identified the article ``a'' in ``... cell remains a pass ...'' as a pronoun.
}.

\vspace{2mm}
\subsection{Potential and Challenges of LOGOS-CA}
\label{sec.logos-ca}

LOGOS-CA retains the same scope of simulation as conventional cellular automata 
while also enabling more flexible descriptions of states and rules. 
If applied to chemical reactions, for example, each cell's state could correspond to a single molecule using notation such as SMILES, 
and various reaction patterns written in natural language could potentially be employed as rules.
Similarly, in material fracture simulations, 
each cell could encapsulate not only simple information such as pressure or velocity, 
but also the possibility of microscopic cracks occurring, expressed as text.
Additionally, for phenomena such as traffic congestion and economic activities, 
there is room to flexibly reflect interactions in which human intentions are involved.

\vspace{2mm}
While this framework holds enormous potential, 
simulations using LOGOS-CA also present notable challenges.

\vspace{2mm}
\subsubsection{LLM Dependency of Results}

\vspace{2mm}
As can be seen from the forest fire example, 
if clear and detailed descriptions are provided, 
LLMs above a certain level of capability are expected to produce the same simulation results, 
and deviations from the rules can be completely monitored.
In contrast, as in the ALife example, when an LLM is given significant discretion, 
the characteristics of the LLM may be strongly reflected in the results.
This is prominently manifested not only in differences in the states themselves—
such as the symbolic state definitions observed only in GPT-5-nano 
and the references to decoration beyond simple text observed only in GPT-5-mini—but also in the update rules.
Particularly in the context of ALife, 
the diversity of states and convergence to fixed solutions are often discussed; 
however, as illustrated in Fig.~\ref{fig:combined_change_timeseries}, these depend significantly on the LLM selected as the engine.

\vspace{2mm}
Therefore, when conducting simulations using LOGOS-CA, 
it is necessary to carefully examine the results according to the degree of flexibility involved.
Specifically, for simulations executed based on explicit rules, 
it suffices to simply check whether the rules are being violated. 
However, if there is any flexibility involved, 
one must consider the possibility that the results may merely be due to the choice of model.
Simulations of phenomena governed by strict rules, 
such as microscopic chemical reactions and macroscopic fracture, fall into the former category, 
whereas ALife and social models involving human decision-making fall into the latter.
Additionally, even for simulations such as chemical reactions, 
if the rules involve probabilities, attention must be paid to whether those probabilities are properly handled by the selected LLM.

\vspace{2mm}
\subsubsection{Cost and Scalability}

LOGOS-CA simulations incur substantial costs in both time and money.
This is because advancing a single step requires LLM processing for every cell.
The costs incurred in the forest fire and ALife experiments are shown in 
Table~\ref{tab:cost_forest_fire} and Table~\ref{tab:cost_alife}, respectively.
The maximum total cost reached \$14 for the forest fire experiment 
and \$91 for the ALife experiment.
The situation regarding both time and cost is expected to improve 
as capable open-source LLMs become available.
In particular, regarding time, since the processing of each cell is 
entirely independent, the wall time per step could ideally be reduced 
to that of processing a single cell, given sufficient computational resources.
Furthermore, depending on the simulation, many cells may have nearly 
identical descriptions to process, in which case sharing and reusing 
KV caches could prove highly effective~\cite{2023KWO, 2024ZHE}.

\begin{table}[t]
\centering
\caption{Forest fire experiment: token usage and estimated cost (USD).}
\label{tab:cost_forest_fire}
\begin{tabular}{lrrrrrr}
\toprule
Model & In tokens & Out tokens & Total tokens & In cost (\$) & Out cost (\$) & Total (\$) \\
\midrule
GPT-4o-mini & 993,933  & 86,373   & 1,080,306 & 0.1491 & 0.0518 & 0.2009 \\
GPT-4o      & 2,030,406& 217,527  & 2,247,933 & 5.0760 & 2.1753 & 7.2513 \\
GPT-5-nano  & 610,171  & 2,326,698& 2,936,869 & 0.0305 & 0.9307 & 0.9612 \\
GPT-5-mini  & 2,021,853& 1,247,415& 3,269,268 & 0.5055 & 2.4948 & 3.0003 \\
GPT-5       & 2,031,348& 1,236,777& 3,268,125 & 2.5392 & 12.3678& 14.9070 \\
\bottomrule
\end{tabular}
\end{table}

\begin{table}[t]
\centering
\caption{ALife experiment: token usage and estimated cost (USD).}
\label{tab:cost_alife}
\begin{tabular}{llrrrrrr}
\toprule
Model & Run & In tokens & Out tokens & Total tokens & In cost (\$) & Out cost (\$) & Total (\$) \\
\midrule
GPT-5-nano & 1 & 28,838,184  & 52,061,002  & 80,899,186  & 1.4419  & 20.8244 & 22.2663 \\
GPT-5-nano & 2 & 31,834,479  & 65,785,520  & 97,619,999  & 1.5917  & 26.3142 & 27.9059 \\
GPT-5-nano & 3 & 30,283,995  & 59,339,570  & 89,623,565  & 1.5142  & 23.7358 & 25.2500 \\
GPT-5-mini & 1 & 84,104,938  & 22,865,166  & 106,970,104 & 21.0262 & 45.7303 & 66.7566 \\
GPT-5-mini & 2 & 108,905,295 & 32,276,422  & 141,181,717 & 27.2263 & 64.5528 & 91.7792 \\
GPT-5-mini & 3 & 57,519,259  & 18,733,485  & 76,252,744  & 14.3798 & 37.4670 & 51.8468 \\
\bottomrule
\end{tabular}
\end{table}

\vspace{2mm}
\section{Conclusion}
\label{sec.conc}

In this study, 
we focused on the rich expressive power of language and proposed LOGOS-CA as a new framework for harnessing it within cellular automata. 

\vspace{2mm}
Experiments targeting the forest fire simulation demonstrated that LOGOS-CA can accurately emulate simple rule-based cellular automata.
However, the accuracy depends significantly on the LLM selected as the engine, and for certain models, the simulation can easily break down.

\vspace{2mm}
In experiments conducted on ALife as a test for more flexible rules, the differences among LLMs were also clearly manifested. 
GPT-5-nano described states and rules in relatively short sentences and independently invented symbolic state definitions on its own.
In contrast, GPT-5-mini described them in relatively long sentences.
For states, it used expressions that ventured into visual characteristics, 
such as ``high-frequency speckle'', ``ultrashort fragments'', and ``ornamentation''. 
Its rules consistently included 
causation and evidence (unanimously occupied / unanimous reinforcement), 
enumeration of denials and constraints (no overwrite / no vacancy ...), 
and necessary conditions and exceptions (only if ... / will clear only if ...).
Additionally, expressions with stacked modifiers such as "overwrite-resistant" and "self-sustaining ..." were also prominent. 

\vspace{2mm}
From these results, it can be seen that in simulations using LOGOS-CA or similar frameworks, 
it is always necessary to consider the choice of model and the degree of flexibility in rules when interpreting the results.
Specifically, for simulations operated under strict rules, 
it suffices to adopt LLMs above a certain level of capability, and verification is also straightforward. 
Conversely, for highly flexible simulations such as ALife, 
one must always pay attention to whether the results are attributable to the choice of LLM.

\vspace{2mm}
LOGOS-CA has the potential for application to 
microscopic behaviors such as chemical reactions and diffusion, 
macroscopic phenomena such as material deformation and fracture, 
and even social phenomena such as traffic congestion, pedestrian flow, and economic activities.
However, the points discussed above must be fully considered when interpreting the results.

\vspace{2mm}
\section{Code Availability}
\label{sec.code}

The source code and data are available at:
\url{https://github.com/A5size/LOGOS-CA}.

\vspace{2mm}
\section{Acknowledgments}

In the preparation of this paper, 
ChatGPT and Claude were used for 
code generation, prompt generation, and English proofreading. 
All outputs were verified by the author, and the author assumes full responsibility for the content of this paper.

\bibliographystyle{unsrtnat}
\bibliography{references.bib}

\newpage

\vspace{2mm}
\appendix
\section{ALife: Temporal Evolution and Clustering}

\subsection{GPT-5-nano}

\begin{figure*}[htbp]
  \centering
  \includegraphics[width=\textwidth]{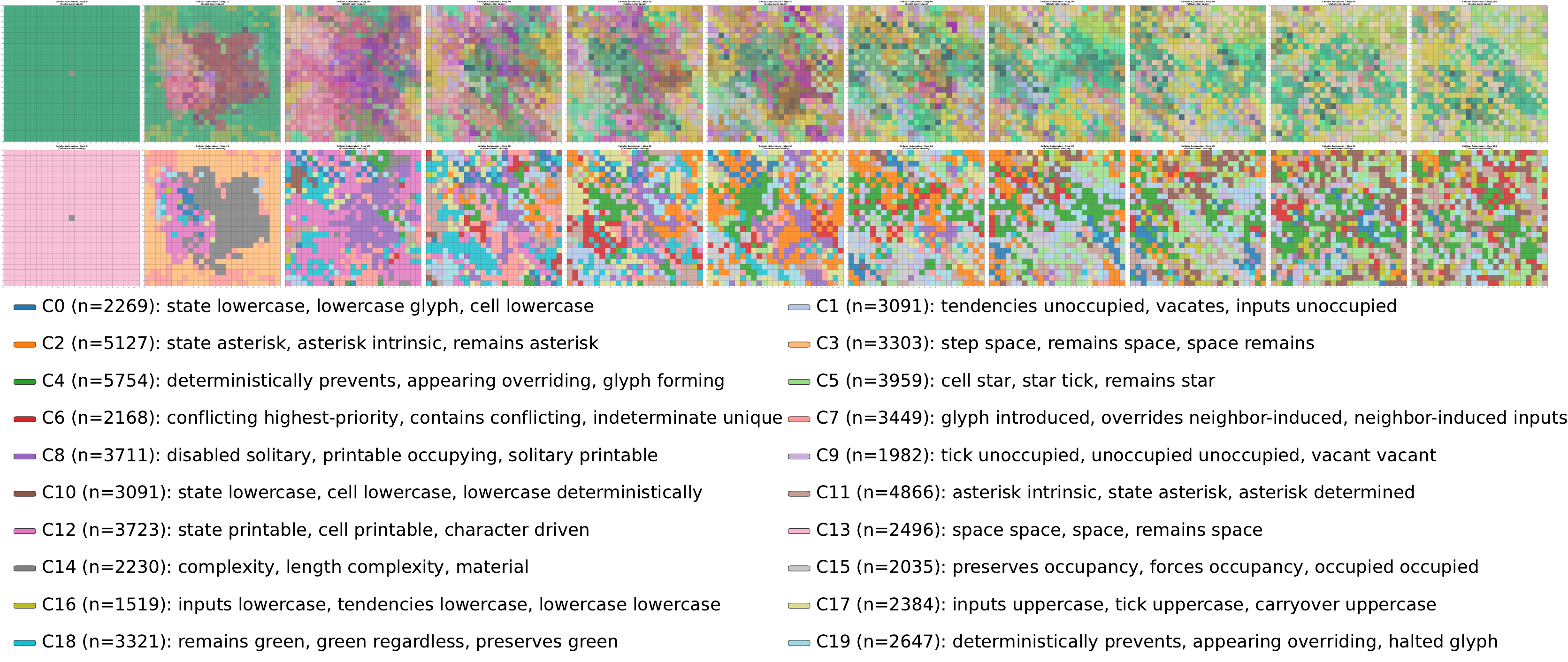}
  \caption{
           Results of Run 2 for GPT-5-nano in the ALife experiment. 
           The top row shows the Global color space and the bottom row shows the cluster-based visualization. 
           See Fig.~\ref{fig:map_gpt-5-nano-01} for details.
           }
  \label{fig:map_gpt-5-nano-02}
\end{figure*}

\begin{figure*}[htbp]
  \centering
  \includegraphics[width=\textwidth]{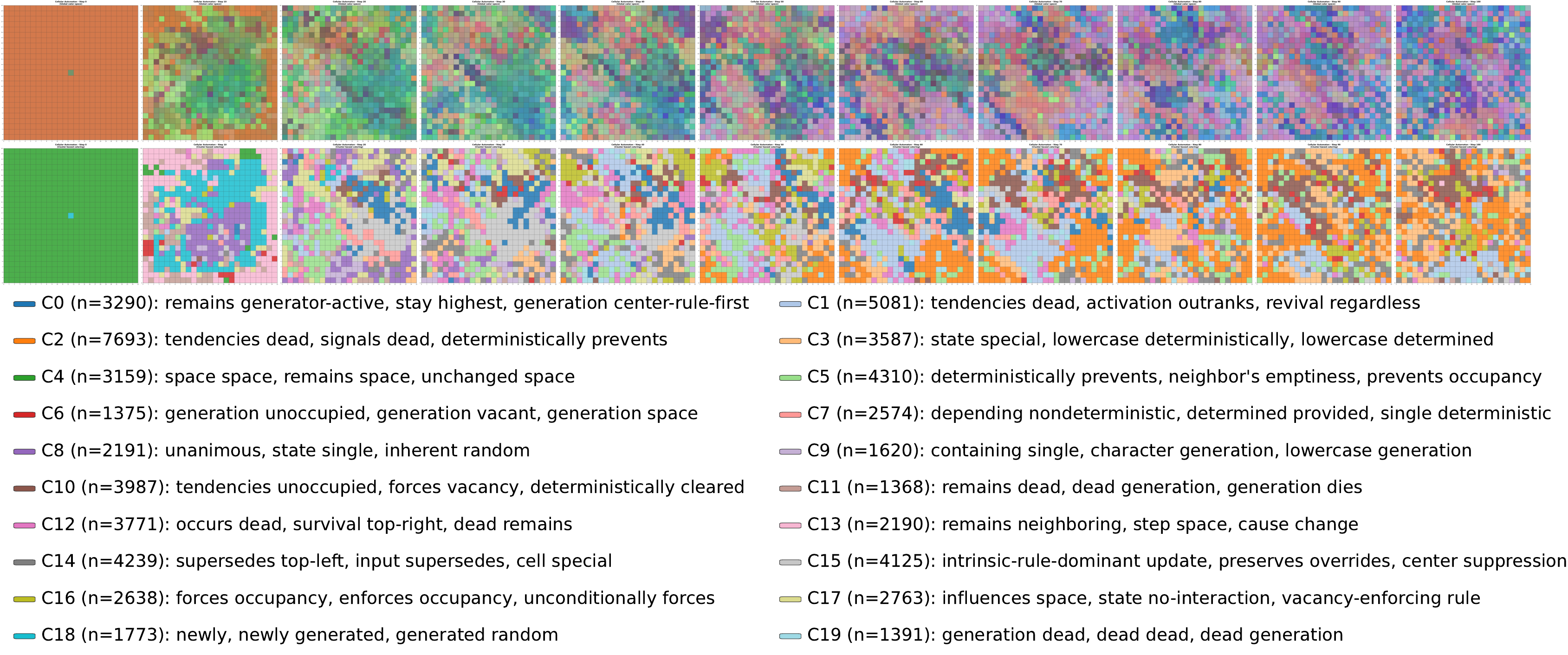}
  \caption{
           Results of Run 3 for GPT-5-nano in the ALife experiment. 
           The top row shows the Global color space and the bottom row shows the cluster-based visualization. 
           See Fig.~\ref{fig:map_gpt-5-nano-01} for details.
           }
  \label{fig:map_gpt-5-nano-03}
\end{figure*}

\newpage

\subsection{GPT-5-mini}

\begin{figure*}[htbp]
  \centering
  \includegraphics[width=\textwidth]{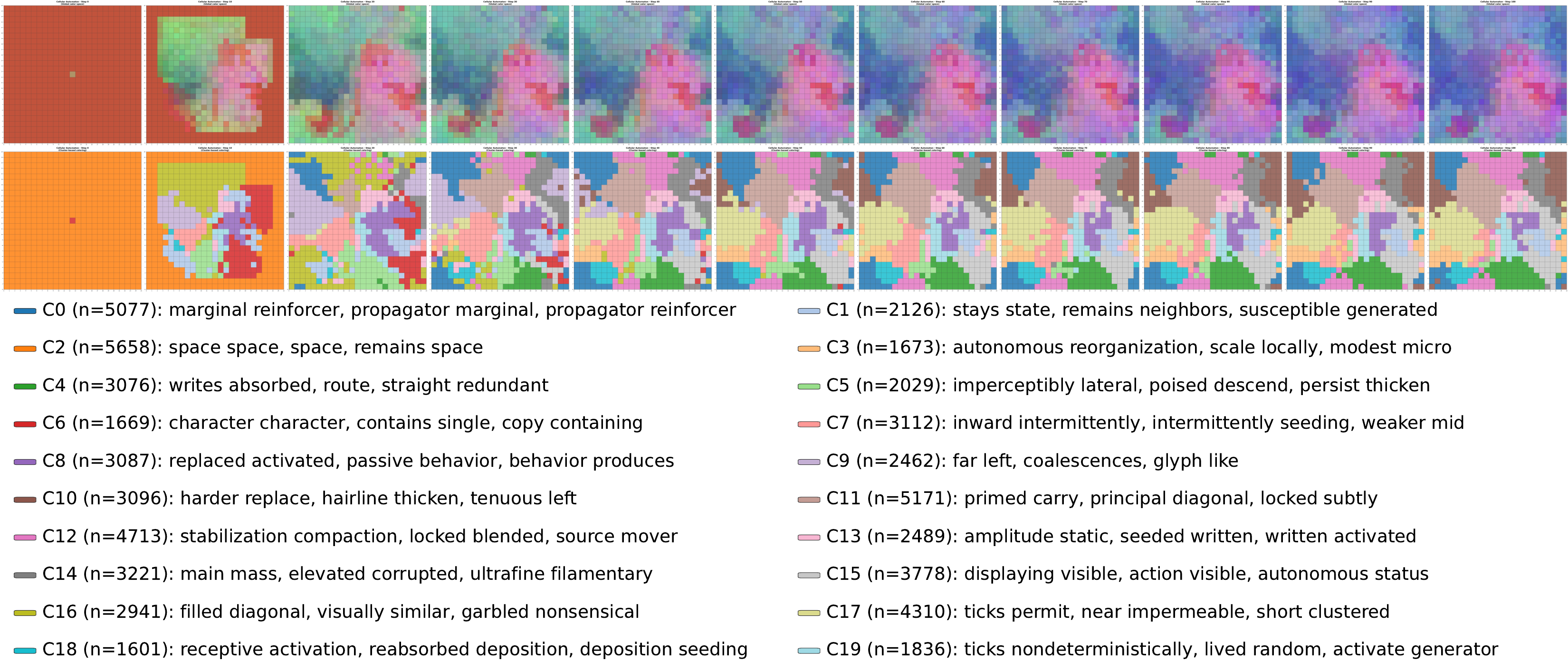}
  \caption{
           Results of Run 2 for GPT-5-mini in the ALife experiment. 
           The top row shows the Global color space and the bottom row shows the cluster-based visualization. 
           See Fig.~\ref{fig:map_gpt-5-mini-01} for details.
          }
  \label{fig:map_gpt-5-mini-02}
\end{figure*}

\begin{figure*}[htbp]
  \centering
  \includegraphics[width=\textwidth]{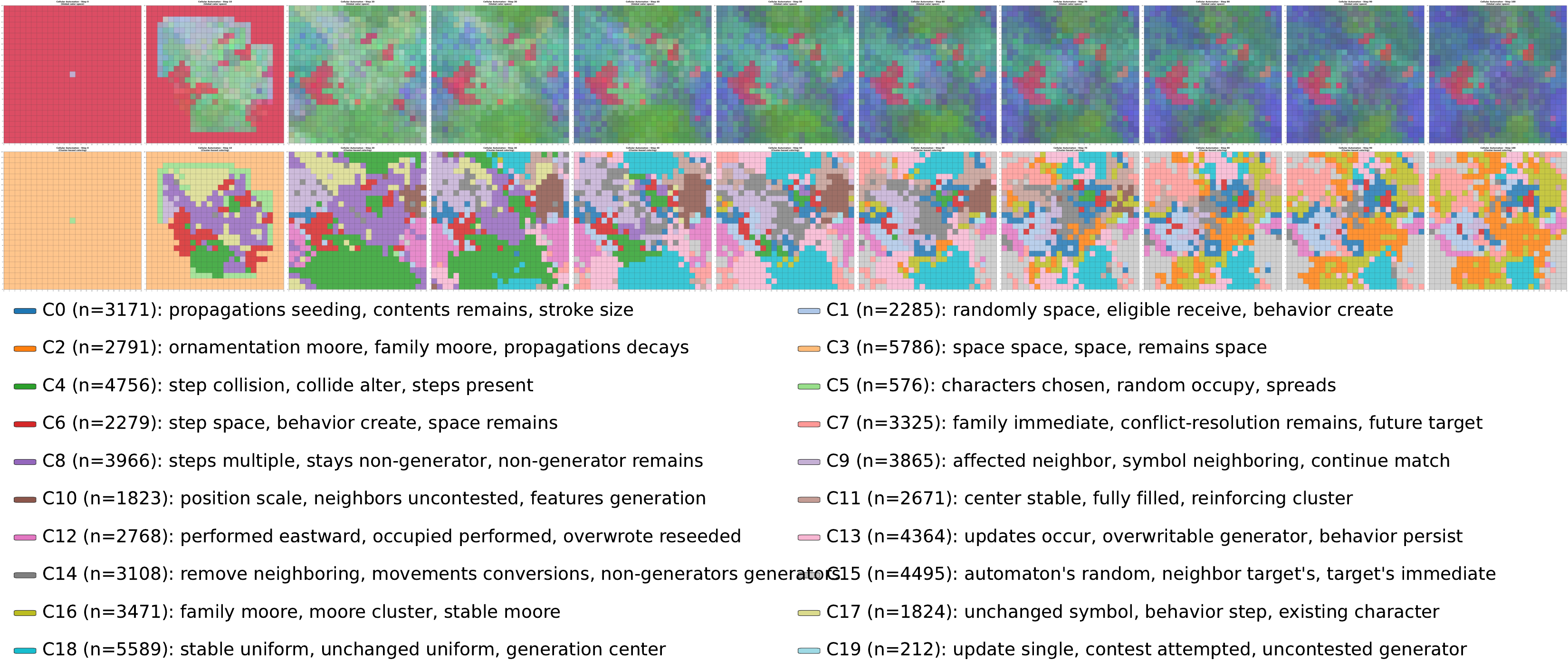}
  \caption{
           Results of Run 3 for GPT-5-mini in the ALife experiment. 
           The top row shows the Global color space and the bottom row shows the cluster-based visualization. 
           See Fig.~\ref{fig:map_gpt-5-mini-01} for details.
          }
  \label{fig:map_gpt-5-mini-03}
\end{figure*}

\end{document}